\documentclass[twocolumn,prb]{revtex4}
\usepackage{amsfonts}
\usepackage[T1]{fontenc}
\usepackage{amsmath,amsbsy,amssymb,graphicx}
\usepackage{times}

\begin{document}

\title{ Exact solutions and topological phase diagram \\
in interacting dimerized Kitaev topological superconductors}
\author{Motohiko Ezawa}

\begin{abstract}
It was recently shown that an interacting Kitaev topological superconductor
model is exactly solvable based on two-step Jordan-Wigner transformations
together with one spin rotation. We generalize this model by including the
dimerization, which is shown also to be exactly solvable. We analytically
determine the topological phase diagram containing seven distinct
phases. It is argued that the system is topological when a
fermionic many-body Majorana zero-energy edge state emerges. It is
intriguing that there are two tetra-critical points, at each of which four
distinct phases touch.
\end{abstract}

\maketitle

\affiliation{Department of Applied Physics, University of Tokyo, Hongo 7-3-1, 113-8656,
Japan}

\textit{Introduction:} Majorana fermions were used for the first time in
condensed matter physics to exactly solve the two-dimensional Ising model by
mapping it to the one-dimensional quantum spin model \cite{Kauf} with the
use of the Jordan-Wigner transformation\cite{JW, Weyl}. Recently, a renewed
interest on Majorana fermions has created one of the most active fields in
the context of topological superconductors\cite{Alicea,Been,Elli}. They are
expected to play a key role in future topological quantum computations\cite{TQC}. 
The Kitaev topological superconductor (KTSC) model is a fundamental
one which hosts Majorana fermions\cite{Kitaev}. It is exactly solvable since
it describes free electrons. An interplay between topology and interaction
is a fascinating subject. There are several works where electron-electron
interaction effects have been investigated\cite{Fidkowski, Loss,
Stoudenmire, Lutchyn, Sela, Cheng, Hassler, Thomale, Manolescu, Chan,
Klassen, Affleck,Miao,Fl}. It is shown \cite{Sela, Hassler, Thomale} that
there is a topological phase transition between a topological superconductor
(TSC) state and trivial charge-density wave (CDW) state at a certain
interaction strength.

The KTSC model is characterized by the three parameters, i.e., the transfer
integral $t$, the superconducting pairing gap $\Delta $ and the chemical
potential $\mu $. The system is topological for $\left\vert \mu \right\vert
<2t$, while it is trivial for $\left\vert \mu \right\vert >2t$. The
interacting KTSC model contains an additional electron-electron interaction $U$. 
It is exactly solvable under the frustration free condition\cite{Katsura}, 
i.e., $\mu =4\sqrt{U^{2}+tU+(t^2-\Delta^2)/4}$. It is also exactly
solvable at the symmetric point\cite{Loss} $\Delta =U=t$ and $\mu =0$.
Recently, this exact solution is extended for $\Delta=t$ and $\mu =0$ with
an arbitrary $U$ by mapping the system to the KTSC model\cite{FuChun} with
the aid of the combination of two-step Jordan-Wigner transformations and one
spin rotation. This method is also applicable to the KTSC
model with disorders\cite{McG}.

In this paper, we generalize the interacting KTSC model by including the
dimerization with parameter $\eta $, $|\eta |\leq 1$. The model is exactly
solvable for the case of $\Delta =t$ and $\mu =0$ with an arbitrary $U$. We
analytically obtain the topological phase diagram in the ($U/t$)-$\eta $
plane, which contains seven distinct phases. The topological
properties of each phase are determined based on the bulk-edge
correspondence. It is argued that the emergence of a fermionic many-body
Majorana zero-energy edge state is a manifestation of the topological
nontriviality of the system. We also discuss the duality relation between
topological phases.

\begin{figure}[t]
\centerline{\includegraphics[width=0.45\textwidth]{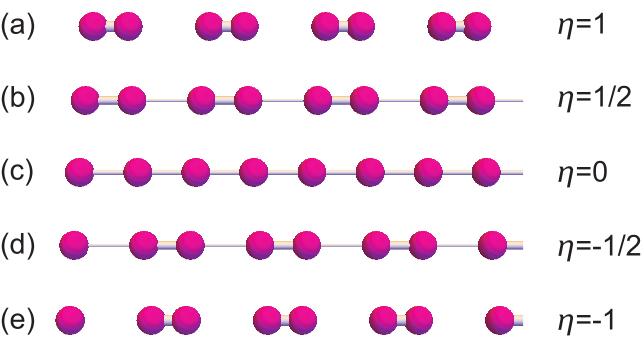}}
\caption{Illustration of a semi-infinite chain of spinless electrons for
various dimerization $\protect\eta $. For the case $\protect\eta =-1$, the
edge site is isolated, leading to the SSH-like zero-energy edge state.}
\label{FigDimer}
\end{figure}

\textit{Hamiltonian:} We consider a one-dimensional chain of spinless
electrons: See Fig.\ref{FigDimer}. The tight-binding model for a hybrid
system comprised of the Kitaev model\cite{Kitaev} and the
Su-Schrieffer-Heager (SSH) model\cite{SSH} together with the
electron-electron interaction is given by 
\begin{align}
H=& -\mu \sum_{j}c_{j}^{\dagger }c_{j}-\sum_{j}t_{j}(c_{j}^{\dagger }c_{j+1}+%
\text{h.c.})  \notag \\
& -\sum_{j}\Delta _{j}(c_{j}^{\dagger }c_{j+1}^{\dagger }+\text{h.c.}) 
\notag \\
& +\sum_{j}U_{j}\left( 2c_{j}^{\dagger }c_{j}-1\right) \left(
2c_{j+1}^{\dagger }c_{j+1}-1\right) ,  \label{KSSH}
\end{align}
with
\begin{eqnarray}
t_{j} &=&t\left\{ 1-\eta \left( -1\right) ^{j}\right\} ,\qquad \Delta
_{j}=\Delta \left\{ 1-\eta \left( -1\right) ^{j}\right\} ,  \notag \\
U_{j} &=&U\left\{ 1-\eta \left( -1\right) ^{j}\right\} ,
\end{eqnarray}
where $\mu $ is the chemical potential, $t$ is the transfer integral, 
and $\Delta $ is the superconducting pairing gap taken to be real. 
Parameters $t_{j}$, $\Delta _{j}$ and $U_{j}$\ are dependent of sites due to the
dimerization $\eta $. We assume $t\geq 0$ without loss of generality, since
the local unitary transformation $c_{j}\rightarrow -i\left( -1\right)
^{j}c_{j}$ interchanges $t$ and $-t$. In addition, we assume $\Delta \geq 0$
since the phase \ transformation $c_{j}\rightarrow ic_{j}$ interchanges $\Delta $ and $-\Delta $. 
Without the interaction this Hamiltonian is reduced to the dimerized KTSC model\cite{Wakatsuki}.

\textit{Jordan-Wigner transformation: }The model with no dimerization is
exactly solvable\cite{FuChun} for the case of $\Delta =t$ and $\mu =0$. We
now show that, even if we include the dimerization $\eta $, it is exactly
solvable for the case of $\Delta =t$ and $\mu =0$. We consider the
Jordan-Wigner transformation\cite{Sela,
Fendley,Hassler,Katsura,Klassen,FuChun}, representing the fermion operators
in terms of the spin operator, such that $c_{i}=K_{i}\sigma _{i}^{-}$ and $c_{i}^{\dagger }=\sigma _{i}^{+}K_{i}^{\dagger }$, 
with $K_{i}=\prod\limits_{j=-M}^{i-1}\left( -\sigma _{j}^{z}\right) $ and $\sigma _{i}^{\pm
}=\sigma _{i}^{x}\pm i\sigma _{i}^{y}$. It follows that
\begin{equation}
\sigma _{j}^{x}\sigma _{j+1}^{x}=c_{j}^{\dagger }c_{j+1}+c_{j+1}^{\dagger
}c_{j}+c_{j}^{\dagger }c_{j+1}^{\dagger }+c_{j+1}c_{j},
\end{equation}
and
\begin{equation}
\sigma _{j}^{z}\sigma _{j+1}^{z}=\left( 2c_{j}^{\dagger }c_{j}-1\right)
\left( 2c_{j+1}^{\dagger }c_{j+1}-1\right) .
\end{equation}
The Hamiltonian is rewritten in terms of the spin operator as
\begin{equation}
H=\sum_{j}(-t_{j}\sigma _{j}^{x}\sigma _{j+1}^{x}+U_{j}\sigma _{j}^{z}\sigma
_{j+1}^{z}),
\end{equation}
which is the XZ spin model with dimerization\cite{Perk}. It is customary to make a spin rotation\cite{LSM,FuChun} by $\pi /2$ around the $x$ axis using the
rotation operator $R=\exp \left[ -i\pi /2\sum_{j}\sigma _{j}^{x}\right] $, and
obtain the XY model with dimerization,
\begin{equation}
H=\sum_{j}(-t_{j}\sigma _{j}^{x}\sigma _{j+1}^{x}+U_{j}\sigma _{j}^{y}\sigma
_{j+1}^{y}).
\end{equation}
We further make the inverse Jordan-Wigner transformation, 
\begin{eqnarray}
\sigma _{i}^{x} &=&\frac{1}{2}\left( f_{i}^{\dagger }+f_{i}\right) \exp 
\left[ i\pi \sum_{i}^{j-1}f_{j}^{\dagger }f_{j}\right] , \\
\sigma _{i}^{y} &=&\frac{1}{2i}\left( f_{i}^{\dagger }-f_{i}\right) \exp 
\left[ i\pi \sum_{i}^{j-1}f_{j}^{\dagger }f_{j}\right] , \\
\sigma _{i}^{z} &=&f_{i}^{\dagger }f_{i}-\frac{1}{2}.
\end{eqnarray}
The Hamiltonian turns out to be
\begin{equation}
H=\sum_{j}[-\left( t_{j}-U_{j}\right) f_{j}^{\dagger }f_{j+1}-\left(
t_{j}+U_{j}\right) f_{j}^{\dagger }f_{j+1}^{\dagger }+\text{h.c.}].
\label{HamilF}
\end{equation}
This is solved explicitly as follows.

\textit{Phase diagram:} 
The system has two sublattices $A$ and $B$
made of the odd and even number sites in the presence of the dimerization.
Indeed, the parameters $t_{j}$, $\Delta _{j}$ and $U_{j}$ in the Hamiltonian
(\ref{KSSH}) are common in each sublattice, i.e., $t_{j}=t_{A(B)}$, $\Delta
_{j}=\Delta _{A(B)}$ and $U_{j}=U_{A(B)}$ for all $j$ belonging to the
sublattice $A(B)$. Introducing the four-component operator $C_{k}^{\dagger
}=(f_{kA}^{\dagger },f_{kB}^{\dagger },f_{-kA},f_{-kB})$, where $A$ and $B$
denote the odd and even number sites, we express the Hamiltonian $H$ in the
Bogoliubov-de Gennes form. We obtain
\begin{equation}
H=\frac{1}{2}\sum_{k}C_{k}^{\dagger }\mathcal{H}\left( k\right) C_{k}
\end{equation}
in the momentum space, with 
\begin{equation}
\mathcal{H}\left( k\right) =
\begin{pmatrix}
0 & z & 0 & w \\ 
z^{\ast } & 0 & -w^{\ast } & 0 \\ 
0 & -w & 0 & -z \\ 
w^{\ast } & 0 & -z^{\ast } & 0
\end{pmatrix}
,\label{FinalHamil}
\end{equation}
where 
\begin{align}
z\left( k\right) & =-(t+U)\left[ \left( 1+\eta \right) +\left( 1-\eta
\right) e^{-ika}\right] ,  \label{EqZ} \\
w\left( k\right) & =-(t-U)\left[ \left( 1+\eta \right) -\left( 1-\eta
\right) e^{-ika}\right] ,  \label{EqW}
\end{align}
and $a$ is the lattice constant. 
Diagonalizing this Hamiltonian we obtain the eigenvalues and the eigenfunctions explicitly.

\begin{figure*}[t]
\centerline{\includegraphics[width=0.96\textwidth]{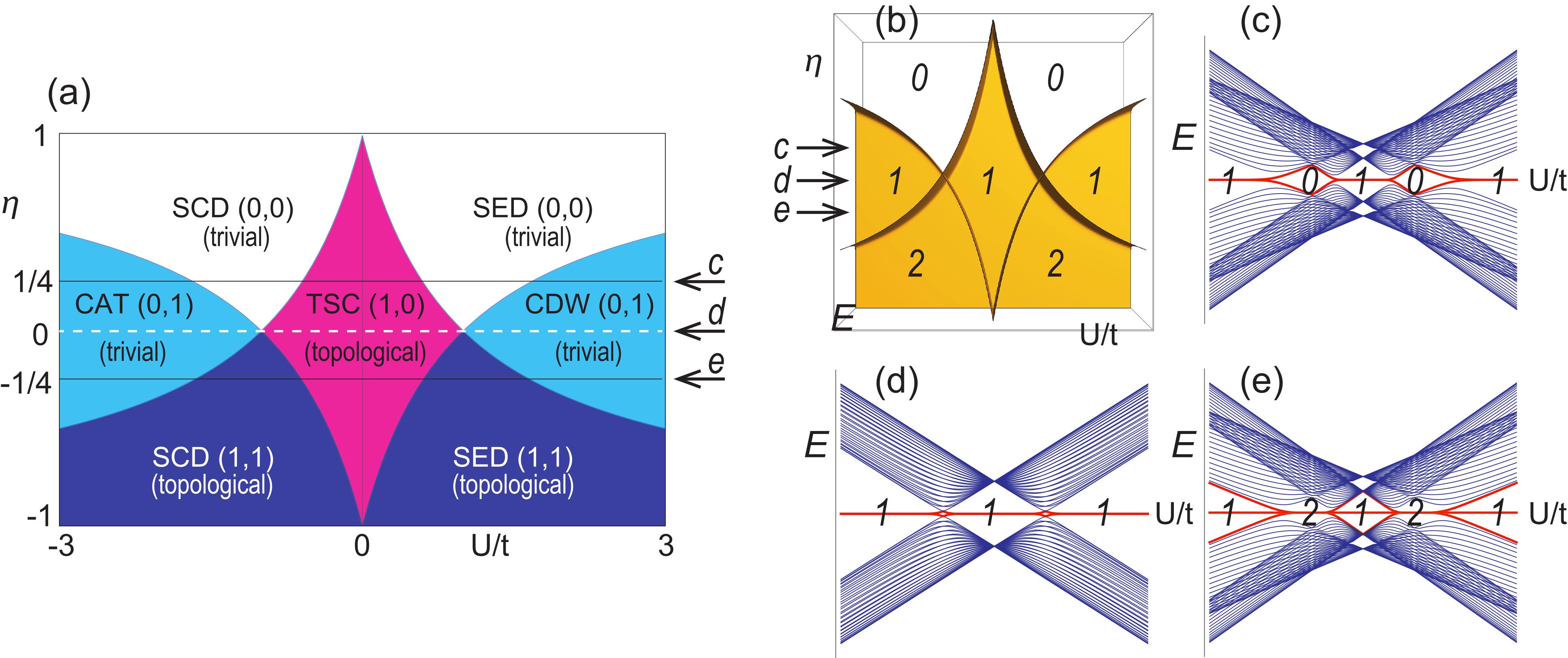}}
\caption{(a) Topological phase diagram on the ($U/t$)-$\protect\eta $ plane.
There are seven distinct phases indexed by ($Q_{I},Q_{II}$), 
where $Q_{\protect\nu }$ is the number of the type-$\protect\nu $ edge states:
Topological-superconductor (TSC), charge-density-wave (CDW), Schr\"{o}dinger-cat (CAT), 
single-electron-dimer (SED) and superconducting-dimer
(SCD) phases. (b) The low-energy edge spectrum on the ($U/t$)-$\protect\eta $
plane. Although there are edge states for the CDW and the CAT phases, they
are bosonic and the systems are trivial. (c)--(e) The energy spectrum of a
finite chain along the $\protect\eta $ line with $\protect\eta =1/4,0,-1/4$,
which is indicated by the arrow labelled by \textit{c, d, e} in (a) and (b).
Edge states are emphasized by red. The horizontal axis represents $U/t$,
upon which the number of the zero-energy edge states is given for a
semi-infinite chain. It is one half of that for a finite chain. }
\label{FigRibbon}
\end{figure*}

In particular, the eigenvalues are 
\begin{equation}
E^{2}\left( k\right) /4=\left( 1\pm \eta \right) ^{2}t^{2}+\left( 1\mp \eta
\right) ^{2}U^{2}-2tU\left( 1-\eta ^{2}\right) \cos k.
\end{equation}
The gap closes for
\begin{equation}
\eta =\pm \left( t-U\right) /\left( t+U\right) ,\quad \pm \left( t+U\right)
/\left( t-U\right) .  \label{Boundary}
\end{equation}
These gap-closing conditions generate the phase boundaries. 
There are seven distinct phases as in Fig.\ref{FigRibbon}(a). 

Our next task is to determine the topological properties of each phase.
However, we cannot discuss the topological properties of the original system with the use of the Jordan-Wigner
transformed operator $f_j$ since it is given by a non-local transformation.
Note that the topological properties are not conserved by such a transformation\cite{Greiter}. 
Nevertheless, it is possible to discuss them by examining the edge state\cite{McG} based on the bulk-edge correspondence
by considering a semi-infinite chain with one edge.

\textit{Majorana edge states:} 
First we show that there are two types of edge states.
We introduce the Majorana representation $\lambda _{j}^{A}=f_{j}^{\dagger }+f_{j}$ and $\lambda
_{j}^{B}=i(f_{j}^{\dagger }-f_{j})$, and rewrite the Hamiltonian (\ref{HamilF}) in the Majorana form,
\begin{equation}
H=i\sum_{j}t_{j}\lambda _{j}^{B}\lambda _{j+1}^{A}+U_{j}\lambda
_{j}^{A}\lambda _{j+1}^{B}.  \label{Hg}
\end{equation}
This is separated into two independent Hamiltonias\cite{McG} as $H=H_{I}+H_{II}$ with 
\begin{align}
H_{I}=& \sum_{j}it_{2j}\phi _{I,j}^{B}\phi _{I,j+1}^{A}+iU_{2j-1}\phi
_{I,j}^{A}\phi _{I,j}^{B}, \\
H_{II}=& \sum_{j}-iU_{2j}\phi _{II,j}^{B}\phi _{II,j+1}^{A}-it_{2j-1}\phi
_{II,j}^{A}\phi _{II,j}^{B},
\end{align}
where we have defined $\phi _{I,j}^{A}=\lambda _{2j-1}^{A}$, $\phi
_{I,j}^{B}=\lambda _{2j}^{B},\phi _{II,j}^{A}=\lambda _{2j-1}^{B}$ and $\phi
_{II,j}^{B}=\lambda _{2j}^{A}$. This decoupling is in essence of the
relation between the XY model in zero field and two independent transverse-field Ising
models\cite{Field,PerkD}.

The Jordan-Wigner transformed Majorana operators $\lambda _{\nu }^{\mu }$
are written in terms of the original Majorana operators as\cite{FuChun,McG} 
\begin{align}
\phi _{I,j}^{A}& =\lambda _{2j-1}^{A}=\gamma
_{1}^{B}\prod_{k=1}^{j-3}[i\gamma _{2k}^{A}\gamma _{2k+1}^{B}](i\gamma
_{j-1}^{A}\gamma _{j}^{A}), \\
\phi _{I,j}^{B}& =\lambda _{2j}^{B}=\gamma _{1}^{B}\prod_{k=1}^{j-2}[i\gamma
_{2k}^{A}\gamma _{2k+1}^{B}](i\gamma _{j}^{A}\gamma _{j}^{B}), \\
\phi _{II,j}^{A}& =\lambda _{2j-1}^{B}=\prod_{k=1}^{j-2}[i\gamma
_{2k-1}^{A}\gamma _{2k}^{B}](i\gamma _{j}^{A}\gamma _{j}^{A}), \\
\phi _{II,j}^{B}& =\lambda _{2j}^{A}=\prod_{k=1}^{j-3}[i\gamma
_{2l-1}^{A}\gamma _{2l}^{B}](i\gamma _{j-1}^{A}\gamma _{j}^{A}),
\end{align}
where we have defined $\gamma _{j}^{A}=c_{j}^{\dagger }+c_{j}$ and $\gamma_{j}^{B}=i(c_{j}^{\dagger }-c_{j})$. 

The zero-energy edge states of a semi-infinite chain 
are constructed\cite{Fendley,Kell} by operating the linear combination of the
operators $Q^{\mu}_{\nu}=\sum_{j\ge0}\alpha_{\nu,j}\phi_{\nu,j}^{\mu}$ to the Fock
vacuum $\left\vert \text{vac}\right\rangle $, with $\mu =A,B$ and $\nu =I,II$, where the coefficients are given by\cite{Fendley} 
\begin{align}
\alpha _{I,j}=-\left( \frac{U(1+\eta )}{t(1-\eta )}\right) ^{j},\quad \alpha
_{II,j} & =-\left( \frac{t(1+\eta )}{U(1-\eta )}\right) ^{j}.  \label{alpha}
\end{align}
The edge is either the $A$ site or the $B$ site, according to which we use the many-body Majorana
operator $Q^{A}_{\nu}$ or $Q^{B}_{\nu}$. 

The condition for the convergence
of the edge state is given by $|\alpha _{\nu ,j}^{\mu }|<1$ for $\nu =I$ and 
$II$. This is actually the condition for the emergence of the zero-energy
edge state in Hamiltonian $H_{\nu}$. For instance, if $|\alpha _{I,j}^{\mu
}|>1$ there is no zero-energy edge state in the Hamiltonian $H_{I}$.
Hence, for each phase of the phase diagram
we calculate (\ref{alpha}) to decide whether $|\alpha _{\nu ,j}^{\mu }|<1$ or not, and
determine the number $Q_{\nu}$ of the type-$\nu$ edge states.
We show the results in the phase diagram as in Fig.\ref{FigRibbon}(a). 

\textit{Topological properties:} The Hamiltonian (\ref{KSSH}) does not
conserve the fermion number $N=\sum_{j}c_{j}^{\dagger }c_{j}$ due to the
superconducting pairing term but conserves the fermion parity\cite{FuChun,McG,Fendley,Turner,Katsura} defined by $Z_{2}^{f}=(-1)^{N}$ since
the superconducting paring term only changes the fermion number by two. It
commutes (anticommutes) with any product of an even (odd) number of fermion
operators. It is rewritten in terms of the Majorana operator as $Z_{2}^{f}=\prod_{j}i\gamma _{j}^{A}\gamma _{j}^{B}$. 
We find that $\{Q_{I}^{\mu },Z_{2}^{f}\}=0$ and $[Q_{II}^{\mu },Z_{2}^{f}]=0$. Hence, the
type-I edge states are fermionic, while the type-II edge states are bosonic.

According to the bulk-edge correspondence, zero-energy edge states
necessarily emerge if the system is topological but the reverse is not true.
On one hand, it follows from (\ref{alpha}) that
the type-I edge state is adiabatically connected to a non-interacting
Majorana zero-energy state as $U\rightarrow 0$, where it is shown to be
topological based on the well-defined topological argument. 
On the other hand, since the type-II edge state disappears except for the $\eta=-1$ case,
it is connected to a trivial state as $U\rightarrow 0$.
It is bosonic, as we have mentioned. Consequently, the system with the type-I edge
state is topological, while that with the type-II edge state is trivial\cite{Fendley}.

\textit{TSC, CDW and CAT phases:} We have found seven distinct phases. 
We investigate their topological and ground-state properties more in
details. First, we focus on the three phases along the $\eta =0$ line in Fig.\ref{FigRibbon}(a). 
They are already known\cite{FuChun} and named the
topological-superconductor (TSC), charge-density-wave (CDW) and Schr\"{o}dinger-cat (CAT) phases. The Schr\"{o}dinger-cat state is a superposition of
two superconducting states with different occupation numbers\cite{FuChun}.
These ground-state properties are extended into the two-dimensional regions
as in Fig.\ref{FigRibbon}(a) for $\eta \neq 0$, and hence we use the same
names also for the two-dimensional phases. We note that the points $(U/t,\eta )=(\pm 1,0)$ are tetra-critical points at which four distinct
phases touch. It has been shown\cite{Sela, Cheng, Hassler, Thomale,FuChun}
that the TSC phase is topological while the CDW and CAT phases are trivial,
which are consistent with the present results.

\textit{Dimer state:} There are four phases which are absent for $\eta =0$.
We name them as the single-electron-dimer (SED) phases, and the
superconducting-dimer (SCD) phases by the ground-state properties.
Their ground-state properties and topological properties are made manifest in the strong dimerization limit $\eta =\pm1$.

For $\eta =1$ the system is separated into independent dimers as in Fig.\ref{FigDimer}(a). 
The Hamiltonian (\ref{KSSH}) reads $H=\sum_{j}H_{2j-1,2j}$, where for instance we have 
\begin{eqnarray}
H_{1,2} &=&-t\left[ c_{1}^{\dagger }c_{2}+c_{2}^{\dagger }c_{1}\right]
-\Delta \left[ c_{1}^{\dagger }c_{2}^{\dagger }+c_{2}c_{1}\right]  \notag \\
&&+U\left( 2c_{1}^{\dagger }c_{1}-1\right) \left( 2c_{2}^{\dagger
}c_{2}-1\right) .
\end{eqnarray}
The diagonalization is straightforward\cite{Katsura}. We note that, since the two-site
Hamiltonian commutes with the fermion parity operator $Z_{2}^{f}$, the
Hilbert space is decomposed into two subspaces containing even or odd
numbers of electrons.

The even subspace is composed of the following two states, 
\begin{equation}
\left\vert 00\right\rangle \equiv \left\vert \text{vac}\right\rangle \text{
and }\left\vert 11\right\rangle \equiv c_{1}^{\dagger }c_{2}^{\dagger
}\left\vert \text{vac}\right\rangle ,
\end{equation}
corresponding to there are no electrons or two electrons. The Hamiltonian in
the basis of $\left\{ \left\vert 00\right\rangle ,\left\vert 11\right\rangle
\right\} $ is given by
\begin{equation}
H=\left( 
\begin{array}{cc}
U & -\Delta \\ 
-\Delta & U
\end{array}
\right) ,
\end{equation}
which yields the energy dispersion $E_{\pm }^{\text{even}}=U\mp \Delta $
with the eigenfunction $\psi _{\pm }^{\text{even}}=\frac{1}{\sqrt{2}}
(\left\vert 00\right\rangle \pm \left\vert 11\right\rangle )$.

The odd subspace is composed of the following two states, 
\begin{equation}
\left\vert 10\right\rangle \equiv c_{1}^{\dagger }\left\vert \text{vac}
\right\rangle \text{ and }\left\vert 01\right\rangle \equiv c_{2}^{\dagger
}\left\vert \text{vac}\right\rangle ,
\end{equation}
corresponding to two one-electron states occupying the first site or the
second site. The Hamiltonian in the basis of $\left\{ \left\vert
10\right\rangle ,\left\vert 01\right\rangle \right\} $ is given by
\begin{equation}
H=\left( 
\begin{array}{cc}
-U & -t \\ 
-t & -U
\end{array}
\right) ,
\end{equation}
which yields the energy dispersion $E_{\pm }^{\text{odd}}=-U\mp t$ with the
eigenfunction $\psi _{\pm }^{\text{odd}}=\frac{1}{\sqrt{2}}(\pm \left\vert
10\right\rangle +\left\vert 01\right\rangle )$.

We may derive the following results. On one hand, when the interaction is
repulsive ($U>0$), the ground state is a symmetric single electron hopping
state $\psi _{+}^{\text{odd}}=\frac{1}{\sqrt{2}}(\left\vert 10\right\rangle
+\left\vert 01\right\rangle )$ with the energy $E_{+}^{\text{odd}}=-U-t$. It
is reasonable to call it the SED state. On the other hand, when the
interaction is attractive ($U<0$), it is a symmetric superconducting state 
$\psi _{+}^{\text{even}}=\frac{1}{\sqrt{2}}(\left\vert 00\right\rangle
+\left\vert 11\right\rangle )$ with the energy $E_{+}^{\text{even}}=U-\Delta 
$. Since the state contains a pair of electrons, it is reasonable to call it
the SCD state. Since there exist no zero-energy states, the system is
topologically trivial.

For $\eta =-1$ the semi-infinite chain system is separated into independent dimers and an extra
single site at the edge, as in Fig.\ref{FigDimer}(e). The analysis of the
dimer parts is precisely the same as in the limit $\eta =1$. The single
electron at the edge plays a key role, since its energy is zero in the
absence of the chemical potential ($\mu =0$). There exist two zero-energy
states; the state $\left\vert 0\right\rangle =\left\vert \text{vac}\right\rangle $ is bosonic, 
while the state $\left\vert 1\right\rangle
=c_{1}^{\dagger }\left\vert \text{vac}\right\rangle $ is fermionic. The
emergence of the fermionic zero-energy state is a manifestation of the
topological nontriviality of the system.

These basic properties remain almost as they are for $\eta \neq \pm 1$. At least in the
region near $\eta =\pm 1$, the ground state is a linear superposition of
individual dimers [Fig.\ref{FigDimer}(b) and (d)]. There are trivial dimer
phases for $\eta >0$, while there are topological dimer phases for $\eta <0$, 
which is differentiated by the emergence of the zero-energy edge state, as
shown in Fig.\ref{FigRibbon}(c)--(e). We find there is no zero-energy state
for $\eta >0$, while there are two zero-energy states per one edge for $\eta
<0$, which are the type-I and the type-II edge states. In the topological
phase, there is an unpaired site at the edge of a semi-infinite chain [Fig.\ref{FigDimer}(d) and (e)], 
which results in the zero-energy edge states. We
note that the topological and trivial phases alter once we take a
half-shifted unit cell, which is a reminiscence of the SSH model\cite{SSH}.

\textit{Duality:} The system has several duality relations\cite{Kauf,Onsager}. 
The system (\ref{HamilF}) with $\eta =0$ is self-dual\cite{FuChun} for $U=t$. We generalize it to the case that $\eta \neq 0$. The
Hamiltonian (\ref{Hg}) is invariant under the duality transformation, $t\leftrightarrow U$ and $\lambda _{j}^{A}\leftrightarrow \lambda _{j}^{B}$.
For $\eta =0$, there is only one transition point for $U/t>0$, where the
self-duality determines the transition point\cite{FuChun} as $U=t$. For $\eta \neq 0$, there are two transition points at $\eta =\pm \left(
t-U\right) /\left( t+U\right) $ corresponding to (\ref{Boundary}), which are
exchanged by the duality transformation.

The Hamiltonian is invariant also under the duality transformation, 
$t\leftrightarrow -U$ and $\lambda _{j}^{A}\leftrightarrow \left( -1\right)
^{j}\lambda _{j}^{B}$, for which a similar argument follows. For $\eta =0$,
there is only one transition point for $U/t<0$, where the self-duality
determines the transition point as $U=-t$. For $\eta \neq 0$, there are two
transition points at $\eta =\pm \left( t+U\right) /\left( t-U\right) $
corresponding to (\ref{Boundary}), which are exchanged by the duality
transformation.

Finally, the Hamiltonian is invariant also under the duality transformation, 
$U\leftrightarrow -U$\ and $\lambda _{j}^{B}\leftrightarrow \left( -1\right)
^{j}\lambda _{j}^{B}$. For $\eta \neq \pm 1$, there are two transition
points, $U/t=\left( 1\pm \eta \right) /\left( 1\mp \eta \right) $ and 
$-\left( 1\pm \eta \right) /\left( 1\mp \eta \right) $\ corresponding to (\ref{Boundary}), 
which are exchanged by the duality transformation (the upper
signs for $\mu >0$ and the lower signs for $\mu <0$). For $\eta =\pm 1$,
there is only one transition point at $U/t=0$, where the system is self-dual.

The author is very much grateful to N. Nagaosa, J.H.H. Perk, H. Katsura and
J. Knolle for helpful discussions on the subject. This work is supported by
the Grants-in-Aid for Scientific Research from MEXT KAKENHI (Grant
Nos.JP17K05490 and JP15H05854). This work is also supported by
CREST, JST (JPMJCR16F1).

\textit{Note added:} After submission of the manuscript, a closely related
paper\cite{YWang} was uploaded in cond-mat/arXiv, where an exact solution on
a similar interacting dimerized topological superconductor is obtained by
using the same method. There is a difference between the two models that the
interaction $U $ is not dimerized in the above paper. Their results are
consistent with ours.

\end{document}